\documentclass[12pt,preprint]{aastex}

\usepackage{natbib}
\usepackage{graphicx}

\begin{document}

\newcommand{\lya}{Lyman-$\alpha$}
\newcommand{\eqw}{\hbox{EW}}
\def\erg{\hbox{erg}}
\def\cm{\hbox{cm}}
\def\sec{\hbox{s}}
\def\f17{f_{17}}
\def\Mpc{\hbox{Mpc}}
\def\Gpc{\hbox{Gpc}}
\def\nm{\hbox{nm}}
\def\km{\hbox{km}}
\def\kms{\hbox{km\,s$^{-1}$}}
\def\Kkms{\hbox{K\,km\,s$^{-1}$}}
\def\yr{\hbox{yr}}
\def\Myr{\hbox{Myr}}
\def\Gyr{\hbox{Gyr}}
\def\deg{\hbox{deg}}
\def\arcsec{\hbox{arcsec}}
\def\microJy{\mu\hbox{Jy}}
\def\zre{z_r}
\def\fesc{f_{\rm esc}}

\def\ergcm2s{\ifmmode {\rm\,erg\,cm^{-2}\,s^{-1}}\else
                ${\rm\,ergs\,cm^{-2}\,s^{-1}}$\fi}
\def\ergsec{\ifmmode {\rm\,erg\,s^{-1}}\else
                ${\rm\,ergs\,s^{-1}}$\fi}
\def\kmsMpc{\ifmmode {\rm\,km\,s^{-1}\,Mpc^{-1}}\else
                ${\rm\,km\,s^{-1}\,Mpc^{-1}}$\fi}
\def\cMpc{\ifmmode {cMpc}\else
                ${cMpc}$\fi}
\def\kpc{{\rm kpc}}
\def\oii{[O{\sc II}] $\lambda$3727}
\def\oiipair{[O{\sc II}] $\lambda \lambda$3726,3729}
\def\oiii{[O{\sc III}] $\lambda$5007}
\def\oiiipair{[O{\sc III}]$\lambda \lambda$4959,5007}
\def\taulya{\tau_{Ly\alpha}}
\def\taubar{\bar{\tau}_{Ly\alpha}}
\def\llya{L_{Ly\alpha}}
\def\ldlya{{\cal L}_{Ly\alpha}}
\def\nbar{\bar{n}}
\def\Msun{M_\odot} 
\def\mdyn{M_{dyn}}
\def\vmax{v_{max}}
\def\sqamin{\Box'}
\def\l43{L_{43}}
\def\ls{{\cal L}_{sym}}
\def\snrat{\ifmmode {\cal S / N}\else
                   ${\cal S / N}$\fi}
\def\siglos{\sigma_{\hbox{los}}}
\def\asf{\alpha_{SF}}
\def\bsf{\beta_{SF}}
\def\SFR{\hbox{SFR}}
\def\rhoeff{\bar{\rho_e}}
\def\mubar{\bar{\mu}}
\def\MsunYr{M_\odot\,\hbox{yr}^{-1}}

\def\sdss0901{S0901}
\def\s0901full{SDSS090122.37+181432.3}
\def\clone{the Clone}
\def\clonefull{SDSS~J120602.09+514229.5}
\def\rcii{r_{[CII]}}
\def\rhcii{r_{[CII],1/2}}
\def\fcii{f_{[CII]}}
\def\vhalf{v_{1/2}}
\def\vrot{\vhalf}
\def\vhhalf{v_{h,1/2}}
\def\vdhalf{v_{d,1/2}}
\def\vchalf{v_{c,1/2}}
\def\Wm2{\hbox{W}\,\hbox{m}^{-2}}
\def\Lsun{L_\odot}

\renewcommand{\thefootnote}{\fnsymbol{footnote}}

\title{
Herschel Extreme Lensing Line Observations: 
Dynamics of two strongly lensed star forming galaxies near redshift $z=2$
\footnote{{\it Herschel} is an ESA space observatory with science 
instruments provided by European-led Principal Investigator consortia 
and with important participation from NASA.}
}

\author{
James E. Rhoads\altaffilmark{1},
Sangeeta Malhotra\altaffilmark{1},
Sahar Allam\altaffilmark{2},
Chris Carilli\altaffilmark{3},
Fran\c{c}oise Combes\altaffilmark{4},
Keely Finkelstein\altaffilmark{5},
Steven Finkelstein\altaffilmark{5},
Brenda Frye\altaffilmark{6},
Maryvonne Gerin\altaffilmark{7},
Pierre Guillard\altaffilmark{8},
Nicole Nesvadba\altaffilmark{8},
Jane Rigby\altaffilmark{9},
Marco Spaans\altaffilmark{10},
Michael A. Strauss\altaffilmark{11}
}

\begin{abstract}
We report on two regularly rotating galaxies at redshift $z \approx 2$,
using high resolution spectra of the bright [CII] $158\mu$m emission line
from the {\it HIFI} instrument on the {\it Herschel} Space
Observatory.  Both \s0901full\  (``\sdss0901'')  and 
\clonefull\ (``\clone'') are strongly lensed and show the double-horned line 
profile that is typical of rotating gas disks.  Using
a parametric disk model to fit the emission line profiles,
we find that \sdss0901\ has a rotation speed $v \sin(i) \approx 120
\pm 7 \,\kms$ and gas velocity dispersion $\sigma_g < 23\, \kms$ ($1\sigma$).
The best fitting model for \clone\ is 
a rotationally supported disk having $v\sin(i) \approx 79 \pm 11 \, \kms$
and $\sigma_g \la 4\,\kms$ ($1\sigma$).
However \clone\ is also consistent with a family of dispersion-dominated models
having $\sigma_g = 92 \pm 20 \,\kms$.
Our results showcase the potential
of the [CII] line as a kinematic probe of high redshift galaxy
dynamics:  [CII] is bright; accessible to heterodyne receivers with
exquisite velocity resolution; and traces dense star-forming
interstellar gas.
Future [CII] line observations with {\it ALMA}\/ would offer
the further advantage of spatial resolution, allowing a clearer
separation between rotation and velocity dispersion.
\end{abstract}

\altaffiltext{1}{School of Earth and Space Exploration,
Arizona State University, Tempe, AZ 85287, USA; 
email James.Rhoads@asu.edu}
\altaffiltext{2}{Space Telescope Science Institute,
Baltimore, MD 21210, USA}
\altaffiltext{3}{National Radio Astronomy Observatory,
Socorro, NM, USA}
\altaffiltext{4}{Observatoire de Paris,
LERMA, CNRS, 61 Av. de l'Observatoire, F-75 014 Paris, France}
\altaffiltext{5}{Department of Astronomy,
University of Texas at Austin,
2515 Speedway, Stop C1400,
Austin, TX 78712, USA}
\altaffiltext{6}{Steward Observatory, University of Arizona, Tucson, AZ, USA}
\altaffiltext{7}{LERMA,24 rue Lhomond, 75231 Paris Cedex 05, France}
\altaffiltext{8}{Institut d'Astrophysique Spatiale,
Centre Universitaire d'Orsay, France}
\altaffiltext{9}{NASA Goddard Space Flight Center, Greenbelt, MD, USA}
\altaffiltext{10}{Kapteyn Astronomical Institute,
University of Groningen, Groningen, The Netherlands}
\altaffiltext{11}{Department of Astrophysical Sciences, 
Princeton University, Peyton Hall, Princeton, NJ 08544, USA}


\section{Introduction}
Typical star-forming galaxies in the nearby universe are supported by 
systematic rotation, with random velocities playing a relatively
minor role.  This indicates a degree of maturity---  the galaxies are old
and established, and their reservoirs of star-forming gas are dominated by
material that was accreted at least one orbital time ago.
However, at the epoch when cosmic star formation reached its peak,
it is likely that galaxies were still accreting gas rapidly.
The relative roles of rotation and velocity dispersion in providing
support to such galaxies could thus be significantly different.
Certainly, the morphologies of star-forming galaxies at $z\sim 2$ 
suggest {\it some} differences in their typical properties: 
regular spirals constituted a much smaller fraction of galaxies
at $z\sim 2$ than they do today.

Dynamical observations of star-forming galaxies at $z\sim 2$ are
challenging.  One successful approach has been to study strong rest-frame
optical emission lines, which are redshifted to near-IR wavelengths
\citep[e.g.][]{ForsterSchreiber09,Lehnert13,Rhoads13b}.  
Spatial resolution is generally limited by seeing to $\ga 0.5''$,
which yields a few resolution elements across the disk of the galaxy.
For a minority of cases it has been possible to achieve
significantly higher spatial resolution, either by using adaptive optics
\citep[e.g.][]{Law09},
or by observing strongly lensed galaxies \citep{Jones10,Frye12,Wuyts12,Jones13}.
Spectral resolution is also usually modest ($R\la 6000$, corresponding
to $\Delta v \ga 50 \kms$), since the faint flux levels of $z\sim 2$
emission lines usually preclude the lower sensitivity of high resolution
spectrographs.  
 
We here adopt a complementary approach to studying the 
dynamical support of $z\sim 2$ star-forming galaxies. We use 
[CII] 158 \micron\ line observations from the {\it Herschel}
Extreme Lensing Line Observations (HELLO) program \citep{Malhotra14}, 
obtained using the {HIFI} instrument \citep{deGraauw10}
on the {\it Herschel} Space Observatory \citep{Pilbratt10}.  
The resulting line profiles are at exquisite spectral resolution,
though entirely unresolved spatially.  In the present paper
we analyze the galaxies \s0901full\ 
\citep[][hereafter \sdss0901]{Diehl09} and
\clonefull, also called ``\clone'' \citep{Lin09}.  
Both galaxies show the double-horned rotation
profile that is characteristic of rotationally supported disks.  The 
shape of this profile contains considerable information on the disks'
dynamical properties.

C$^+$ is the main coolant of neutral gas in galaxies.  Typically, 0.3\% of
far-infrared luminosity emerges in the [CII] 158 $\mu$m line.  Thus,
it can be an excellent tracer of ISM kinematics.  The question then is
where the [CII] emission comes from. [CII] can trace neutral gas, since
carbon has a lower ionization potential than hydrogen.
The brightest [CII] emission in a galaxy comes from photon-dominated
regions (PDRs).  These are surface layers of dense molecular clouds, where
UV light from nearby star formation is dissociating molecular gas.
[CII] probes primarily a density range from a few $\cm^{-3}$, which
is required to excite the transition, up to the critical density of 
$\sim 10^{3.5} \cm^{-3}$.
The integrated [CII] emission from whole galaxies includes 
potentially important contributions from diffuse ionized gas
\citep{Bennett94}
and also from diffuse HI, although the fractions are debated
\citep[e.g.][]{Madden93,Heiles94,Contursi02}. 
The radial distribution of [CII] in nearby galaxies generally follows
CO \citep[e.g.][for M33 and the Milky Way,
respectively]{Kramer13,Pineda13}, although [CII] emission has
also been observed in outflows \citep{Contursi13}
and in turbulently heated gas in nearby radio-galaxies \citep{Guillard13}.
The galaxies studied in this paper have typical [CII]/FIR ratios
\citep{Malhotra14}, and do not show the [CII] deficiency seen in some
luminous nearby systems \citep[e.g.][]{Malhotra97,Malhotra01,Carilli13}.

We describe our sample and observations in section~\ref{sec:samp_dat}.
We discuss our Markov Chain Monte Carlo spectral line fitting 
methods and results in section~\ref{sec:fitting}, and compare them
to H$\alpha$ observations of the same objects in section~\ref{sec:halpha}.
We close with a discussion of implications and future directions
in section~\ref{sec:discuss}.
Throughout the paper, we adopt a $\Lambda$-CDM ``concordance
cosmology''  with $\Omega_M = 0.27$, $\Omega_\Lambda = 0.73$, and
$H_0 = 71 \kmsMpc$.

\section{The Sample and the Observations}
\label{sec:samp_dat}
\subsection{Sample Selection and Properties}
The full HELLO sample consists of 15 strongly lensed galaxies,
at redshifts $1.0 < z < 3.0$.  The sample was selected to span
a range of both star formation rate and redshift.
All HELLO targets were observed using  the HIFI
Wide Band Spectrometer backend.
We describe the HIFI observations of the full sample 
in \citet{Malhotra14}.  Here, we focus our attention on the two objects
(\sdss0901 and \clone) that show distinct double peaks in their line
profiles.  We summarize the key observational information for
these two objects below. 
Both were identified as candidate lens systems through a search for
blue objects near luminous red galaxies (LRGs) in {\it Sloan} Digital Sky
Survey imaging, followed by visual inspection for arc-like morphology
\citep{Diehl09, Lin09}.

\sdss0901\ was confirmed as a strongly lensed galaxy at redshift
$z=2.2558$ by \citet{Diehl09}.  The lens is a galaxy group, dominated
by a luminous red galaxy at $z=0.3459$ \citep{Diehl09}, and
containing another half dozen smaller galaxies.
The lensed galaxy appears as a $17''$ long
arc, oriented primarily north-south, and located east of the lensing
galaxy.  \citet{Diehl09} identify two main components of the arc
(labeled ``b'' and ``d'' in their published image), and measure its
Einstein radius as $7.7'' \pm 1.1''$.  HST imaging (from program
11602, PI S. Allam) shows considerable substructure (as usual for star
forming galaxies observed in the rest-frame UV), as well as a possible
counterimage $4''$ west of the LRG, and a foreground galaxy superposed
on arc component ``b.''  \citet{Hainline09} published the lensed
galaxy's rest-frame optical spectrum, which shows prominent emission
lines of H$\alpha$ and [NII], as well as weaker lines of [OIII],
H$\beta$, [OII], and [NeIII].   \citet{Hainline09} note that
[NII]/H$\alpha \ga 0.65$, implying the presence of an AGN.  On the other hand,
the object's mid-infrared spectrum shows strong PAH emission features,
suggesting that the dust heating is dominated by star formation rather
than by the AGN \citep{Fadely10}.
The gravitational lensing amplification of \sdss0901\ 
appears not to be formally published.  Available estimates
range from amplification $\mu \sim 8$ (\citet{Fadely10}, quoting from work in
progress by A.~West et al), to $\mu \sim 6$ (Buckley-Geer 2013, private
communication), while inspection of the HST imaging suggests that
larger amplifications are plausible.  
Where needed, we use a fiducial value $\mu=6$, 
but we retain $\mu$ in our equations to show explicitly how the results
will scale for other $\mu$.  Apart from the inferred star formation rate
of \sdss0901, our primary conclusions in the present paper are
essentially independent of the lensing amplification.

The Clone (\clonefull) was first identified as a strongly lensed
galaxy at redshift $z=2.00$ by \citet{Lin09}.  The lens is again a
group, here with three identified members, dominated by an LRG at
$z=0.422$ \citep{Lin09}.  The arc has three primary components which
together span about $120^\circ$, plus a small counterimage.
\citet{Lin09} labeled the components of the main arc as A1, A2, A3,
and the counterimage as A4.  They modeled the lensing geometry based
on ground-based imaging, finding an Einstein radius $3.82''$, and an
amplification factor $\mu = 27\pm 1$.  The lensed galaxy's rest-frame
optical spectrum \citep{Hainline09} is that of a star-forming galaxy,
with no evidence for AGN activity.  Detected lines include H$\alpha$,
H$\beta$, [OII], [OIII], and [SII].
The Clone's mid-IR spectrum is again that
of a normal star-forming galaxy, with prominent PAH emission features
\citep{Fadely10}.  More recent lens modeling, using higher resolution
{\it Hubble Space Telescope} images \citep{Jones10}, gives a refined
amplification $\mu = 28.1 \pm 1.4$. (This is the total amplification
for all lensed images combined, as appropriate for our spatially 
unresolved HIFI observation.)

\citet{Hainline09} also estimated the star formation rate for \clone.
They scaled from the H$\alpha$ flux, assumed a \citet{Chabrier03}
initial mass function, and corrected for gravitational lensing
amplification ($\mu$, see below).  They obtained $\SFR = 32 \MsunYr$
before dust correction, and $\SFR=64$ to $75 \MsunYr$ after dust
correction (depending on the adopted reddening indicator).
They did not estimate a star formation rate for \sdss0901, given
their concerns about the possible role of an AGN.  We can obtain
a crude upper limit on the SFR of \sdss0901\ by neglecting AGN
contributions to H$\alpha$ and applying the same scalings 
used for \clone.  This yields $\sim 135 \MsunYr \times (6/\mu)$,
prior to any dust correction.

\subsection{{\it Herschel} observations}
We observed the {\it HELLO} sample 
using the HIFI Wide-Band Spectrometer (WBS) with 1.10 MHz
resolution.
We used dual beam switching with fast
chopping, to achieve the best possible baselines in our spectra.  For
\sdss0901, we used mixer band 1b and a total observing time of 3000s,
of which 885s was on-source integration time.  For \clone, we used
mixer band 1b with a total observing time of 6200s, of which 1977s was
on-source integration time.  Our data reduction--- which we discuss in
greater detail in \citet{Malhotra14}--- followed mostly standard HIFI
procedures.  The last step of this analysis was to fit and subtract a
baseline from each spectrum.  This means that the spectra used in the
present paper are ``line-only'' spectra, insensitive to the presence
of continuum emission.

The HIFI WBS offers an intrinsic velocity resolution of $<0.6 \kms$ at
the frequency of the redshifted [CII] line in our targets.  However,
each 1.10 MHz resolution element has a signal-to-noise ratio $s/n < 1$
in our data.  Our ability to distinguish fine detail in the spectral
line profiles is therefore effectively limited by sensitivity, since we
can detect features only when we combine enough frequency channels
to get $s/n$ of a few in the feature.

\begin{figure}
\epsscale{0.85}
\plotone{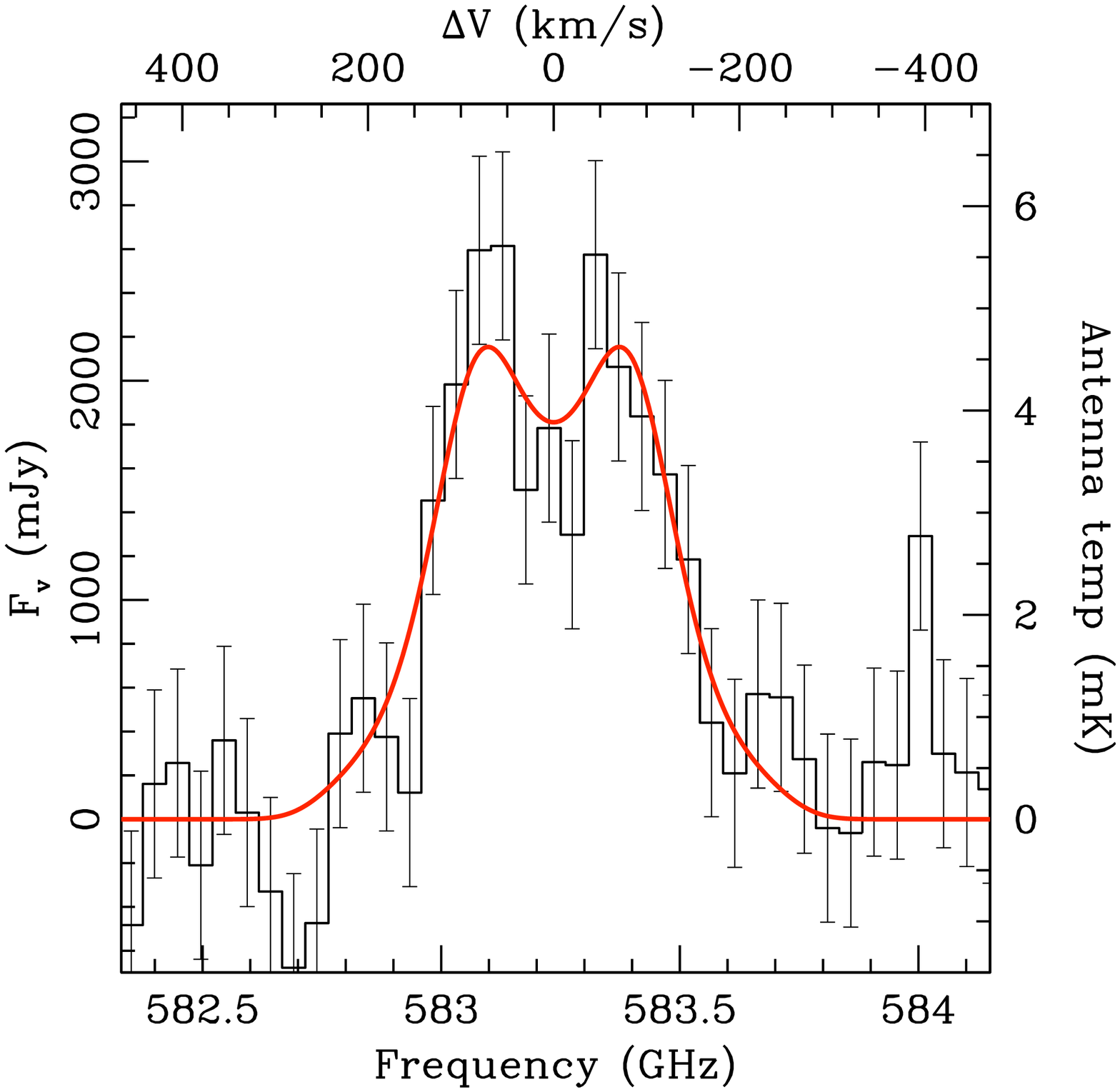}
\caption{HIFI spectrum of the [CII] 158 $\mu$m line from
\s0901full, binned to 25 \kms\ resolution (black histogram).  
Plotted error bars are derived empirically from the RMS 
of the spectrum, measured at line-free frequencies.  
The overplotted model is a centrally concentrated disk model, with
$\vdhalf = 121 \kms$, $\sigma_g = 29 \kms$, and a negligible halo
component.  The resulting ratio of circular speed to velocity
dispersion is 4.2, and at the 95\% confidence level 
the ratio is $> 2.7$.}
\label{fig:s0901_spec}
\end{figure}

\begin{figure}
\epsscale{0.85}
\plotone{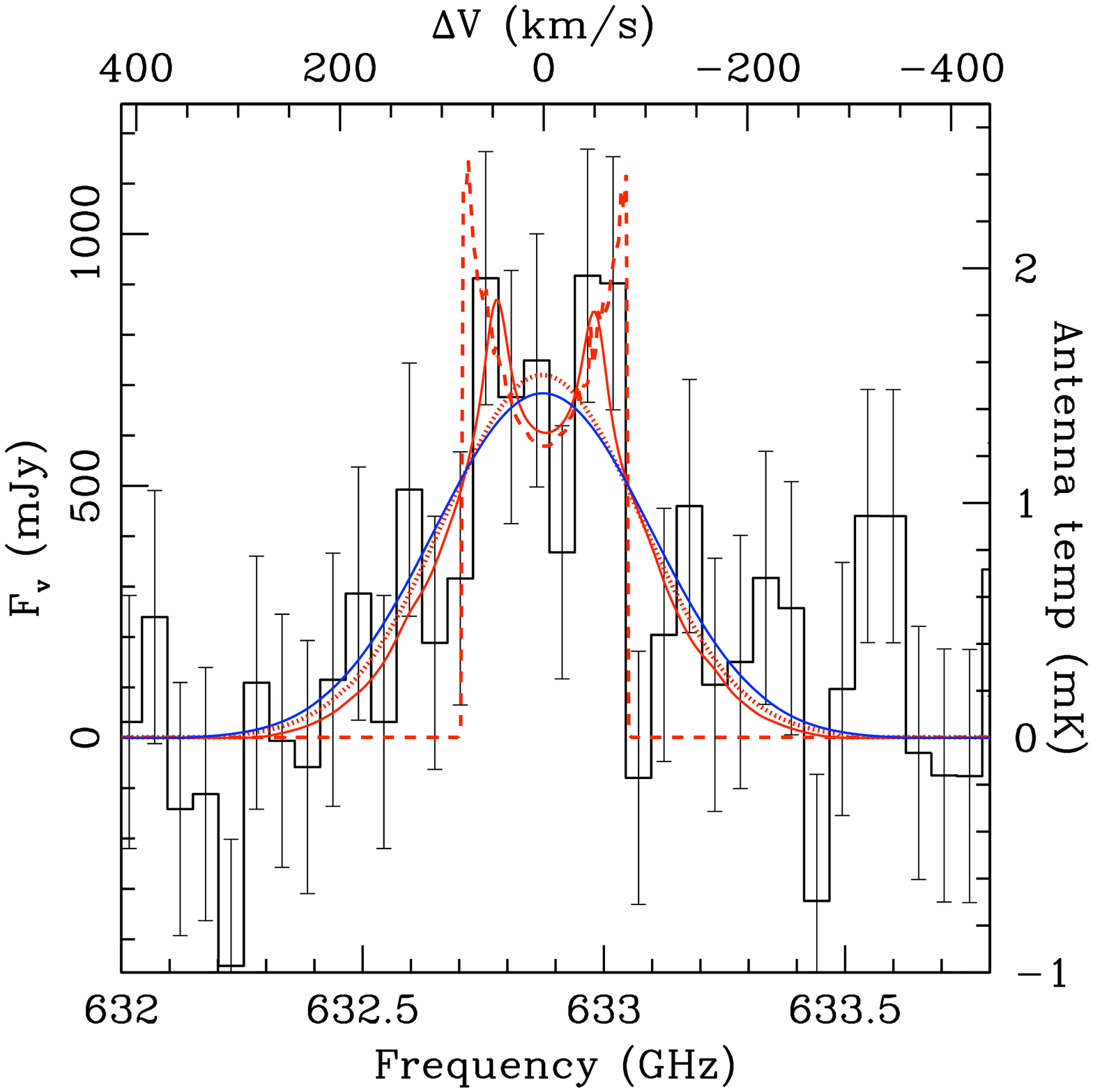}
\caption{Like figure~\ref{fig:s0901_spec}, but for
\clonefull\ (\clone).  Here, the signal-to-noise ratio
is lower, and a range of models provide
acceptable fits to the data.  The modeling procedures
are described in section~\ref{sec:fitting}.  
We plot four representative
cases.  The dashed red line shows the model with the highest 
likelihood among our MCMC simulation results, a ``pure'' 
double-horned line profile where most of the emitting gas is
on the flat part of the rotation curve,
and the velocity dispersion is very small.
If we consider only models with $\sigma_g > 7 \kms$ (motivated
by typical motions within thin disks), the best fit (shown as the 
solid red line) 
remains double-horned, but with very extended wings,
here driven by a rotation curve that continues rising beyond the
spatial extent of the emitting gas.  Finally, we plot 
two dispersion-dominated models.  The first (red dotted line) 
is effectively a pure Gaussian, with $\sigma_g=98 \kms$ and 
negligible rotation.  The second (blue solid
line) matches the results of \citet{Jones13}, from spatially
resolved near-IR spectra, having $\vdhalf=70\kms$,
$x_{ds} = 1.1$, and $\sigma_g = 94 \kms$.  The ratio of
circular speed to velocity dispersion spans a wide range 
for these and other acceptable models of \clone's [CII] emission 
(see figure~\ref{fig:vr_sig}).}
\label{fig:clone_spec}
\end{figure}

We plot the [CII] spectrum of \sdss0901\ in figure~\ref{fig:s0901_spec}
and of \clone\ in figure~\ref{fig:clone_spec}, both at 25 \kms\ 
binning.  \sdss0901\ has a flux of $1.3 \Kkms$ ($1.2\times 10^{-17} \Wm2$),
and yields a signal-to-noise ratio $s/n \ga 4$ per bin for 11 bins above
half-maximum intensity.  \clone\ is fainter, with a flux of $0.3\Kkms$
($3.2\times 10^{-17} \Wm2$), and yields $s/n \sim 3$ per bin for
six bins above half-maximum intensity.  We discuss these
flux measurements further in \citet{Malhotra14},
in the context of the full HELLO sample.
In addition to the spectra, we plot model line profiles from our
fitting procedures in figures~\ref{fig:s0901_spec}
and~\ref{fig:clone_spec}.

\section{Emission line profile fitting}
\label{sec:fitting}
The observed [CII] emission line profiles depend on 
the rotation curves of the lensed galaxies.  They also depend 
on the spatial distribution of the emitting gas, which
determines how the different parts of the rotation curve
are weighted in forming the line profile.  
To model the line profiles, we adopt commonly used fitting functions
for the distributions of both emitting material 
and of mass (which in turn determines the 
rotation curve).

For the radial distribution of [CII] surface brightness, we have
assumed an exponential disk model ($\Sigma_{[CII]} = \Sigma_0
\exp(-r/\rcii)$), which is able to produce models that fit the
observed lines well (see below).  Of course, this does not mean that
the gas distribution {\it must} follow an exponential distribution.
Different tracers of galaxies' interstellar gas often follow different
distributions.  Neutral hydrogen (HI) tends to have a spatially
extended distribution.  Molecular gas can be centrally concentrated
(as in many starbursts and ULIRGs), or can show ring features, as in
both our own Galaxy and M31.  H$\alpha$ comes primarily from dense,
ionized gas, and often shows complex and patchy distributions that
trace the distribution of young star forming regions.

A double-horned line profile basically requires that most of the
emitting material be in significant motion with respect to the
center of mass.  Given that galactic rotation curves rise steeply
in the central regions and then flatten out, there are two ``usual''
ways to achieve this.  First, the gas can be in an extended disk, with
most of the mass on the flat part of the rotation curve.  This is
the common case for HI \citep{Springob05}.
Second, the gas can be in a more compact ring-like configuration, with
little material to emit in the regions where rotation is slow.  This
has been observed, e.g., for CO line emission in NGC 759
\citep[][]{Wiklind97}.   Since [CII] 158 $\mu$m emission comes in
some combination from the diffuse warm neutral medium,
photon-dominated regions at molecular cloud surfaces, {\it and} 
ionized regions, we do not expect a simple analogy with other main
ISM tracers.  We begin by considering the extended disk possibility in detail,
and then examine how our conclusions would be affected for
the case of a [CII] ring.

We model the rotation curve using a disk + halo model for the
gravitating mass.  The halo has mass volume density $\rho \propto
(1+r/r_h)^{-2}$ and an asymptotic rotation speed $v_h$.  The disk
component has mass surface density $\Sigma \propto \exp(-r/r_d)$ and a
peak rotation speed $v_d$.  We can relate $v_d$ to other disk
parameters as $v_d \approx 0.5463 \sqrt{G M_{disk} / r_{d}}$, where
$M_{disk}$ is the total mass of the exponential disk.  $v_d$
corresponds to the circular speed for the disk mass interior to
$r=1.7933 r_{d}$.\footnote{Rephrased, the rotation curve for a pure
  exponential disk peaks near 1.8 scale lengths.}
We model the line-of-sight gas velocity dispersion (due to thermal
{\it and}\/ turbulent motions) as a Gaussian with RMS width $\sigma_g$,
which is convolved with the line profile produced by ordered rotation.

Because our targets are spatially unresolved in HIFI data,
our predicted line profile is insensitive to a global rescaling of 
all radii by a common, constant factor.  We therefore fix the scale 
length $\rcii$ of the surface brightness distribution throughout the modeling,
and rewrite the scale lengths of the disk and halo in dimensionless
form as $x_{d} \equiv r_{d}  / \rcii$ and $x_{h} \equiv r_{h} / \rcii$.
Similarly, the line profile depends on the line-of-sight velocity
$v \sin(i)$ (where $i$ is the inclination angle of the disk) 
but cannot distinguish changes in $i$ from changes in $v$.

The line shape is then determined by a total of five free parameters:
Three velocities ($v_d$, $v_h$, and $\sigma_g$), and two scale length
ratios ($x_{d}$ and $x_{h}$).  We need two further parameters, the
redshift $z$ and total line flux $\fcii$, to completely specify
the line.

Before fitting the disk+halo rotation curve
models, we recast the three velocity parameters in a new form 
that reduces correlated uncertainties in the parameter space. 
Since the observed line profile is determined by gas kinematics
{\it where the light is emitted}, we define the rotational speed
contributions due to the halo and disk, $\vhhalf$ and $\vdhalf$ 
measured at the half-light radius of the [CII] disk (which is 
$\rhcii = 1.66\rcii$).  We then define the {\it total ordered rotation speed}
$\vchalf$ at the half-light radius by
\begin{equation}
\vchalf^2 = \vhhalf^2 + \vdhalf^2 ~~.
\end{equation}
We further define
\begin{equation}
\vhalf^2 = \vchalf^2 + \sigma_g^2 = \vhhalf^2 + \vdhalf^2 + \sigma_g^2 ~~.
\end{equation}
The parameter $\vhalf$ now contains essentially all the information required
to determine the [CII] line width. The relative importance of random and ordered
motions is captured by the velocity ratio 
\begin{equation}
y_{\sigma c} = \sigma_g / \vchalf
\end{equation}
while the relative importance of disk and halo terms in the potential
is captured by 
\begin{equation}
y_{dh} = \vdhalf / \vhhalf ~~.
\end{equation}
The transformation is essentially from Cartesian coordinates
to spherical polar coordinates for the three-space of velocity
parameters, with the $y$ parameters corresponding to (co)tangents
of two angles. 

The final set of parameters for fitting then becomes 
$\left\{\vhalf, y_{\sigma c}, y_{dh}, x_{d}, x_{h}, z, \fcii \right\}$.
By recasting the three velocity parameters,
we eliminate major degeneracies between $v_d$, $v_h$, and $\sigma_g$, 
all of which have the same lowest-order effect of increasing the 
model line width.  Now, $\vhalf$ determines line width, while
the remaining parameters control the shape of the line--- whether it
is singly or doubly peaked; the splitting of the peaks and depth of the
gap between them; and the shape of the high-velocity line wings.

We fitted the rotation curve models to the data using
a Markov Chain Monte Carlo (MCMC) approach (more specifically, the
Metropolis-Hastings algorithm).  In brief, the method consists
of a chain of random steps in our parameter space.  In each step,
the algorithm first identifies a trial solution that is offset
from the current solution by some randomly determined amount.
The likelihood of the trial solution is compared to that of the
current solution.  The trial solution is adopted as the new
solution whenever its likelihood is higher.  If its likelihood
is lower, it may still be adopted as the new solution,
with a probability equal to the ratio of likelihoods.  After 
many steps, the distribution of accepted points
provides an unbiased estimator for the probability distribution 
of the model parameters.  Since consecutive steps of the chain are
highly correlated, we record the state of the chain every 
100 steps,
so that the autocorrelation among the recorded points in parameter space
is weak.

\begin{figure}
\plottwo{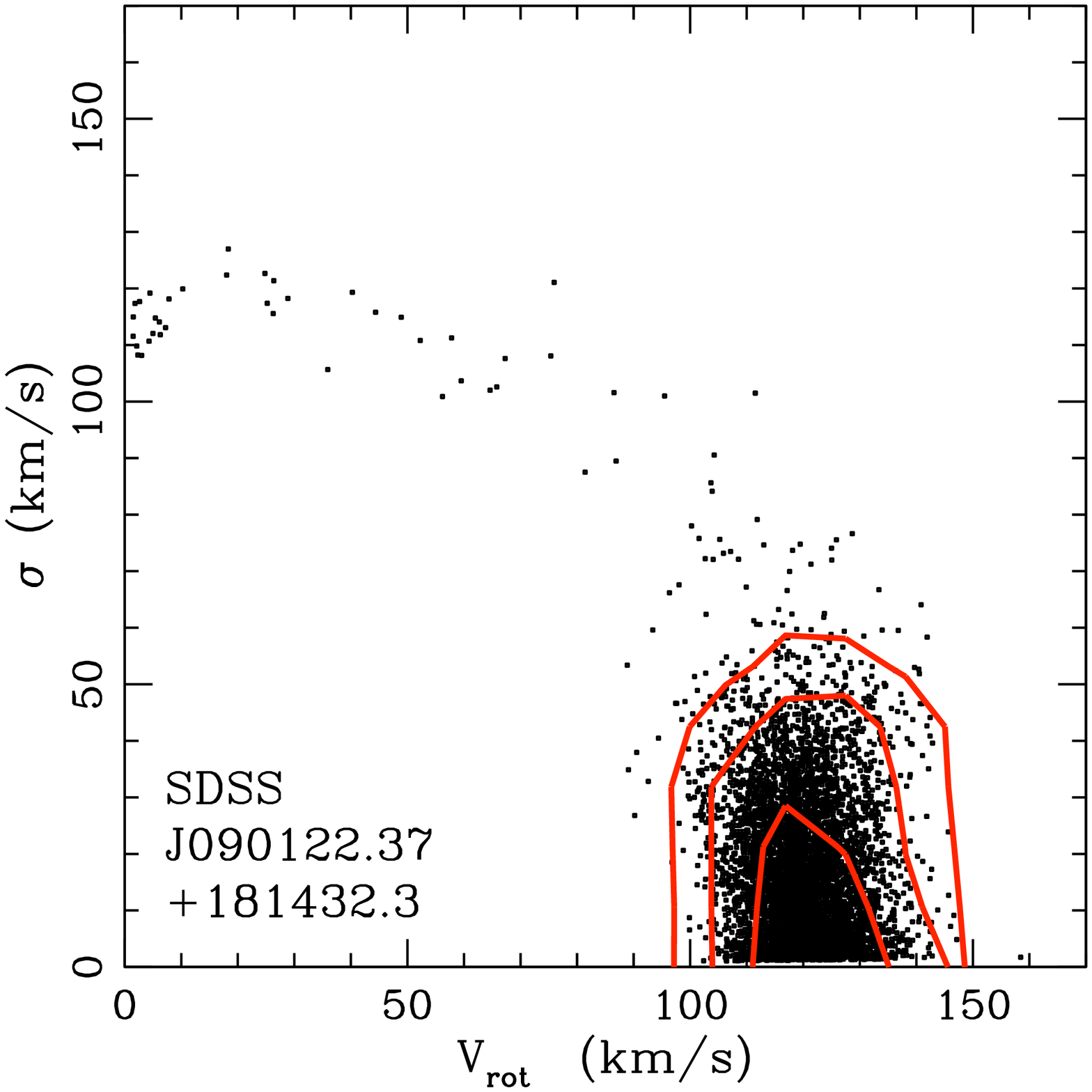}{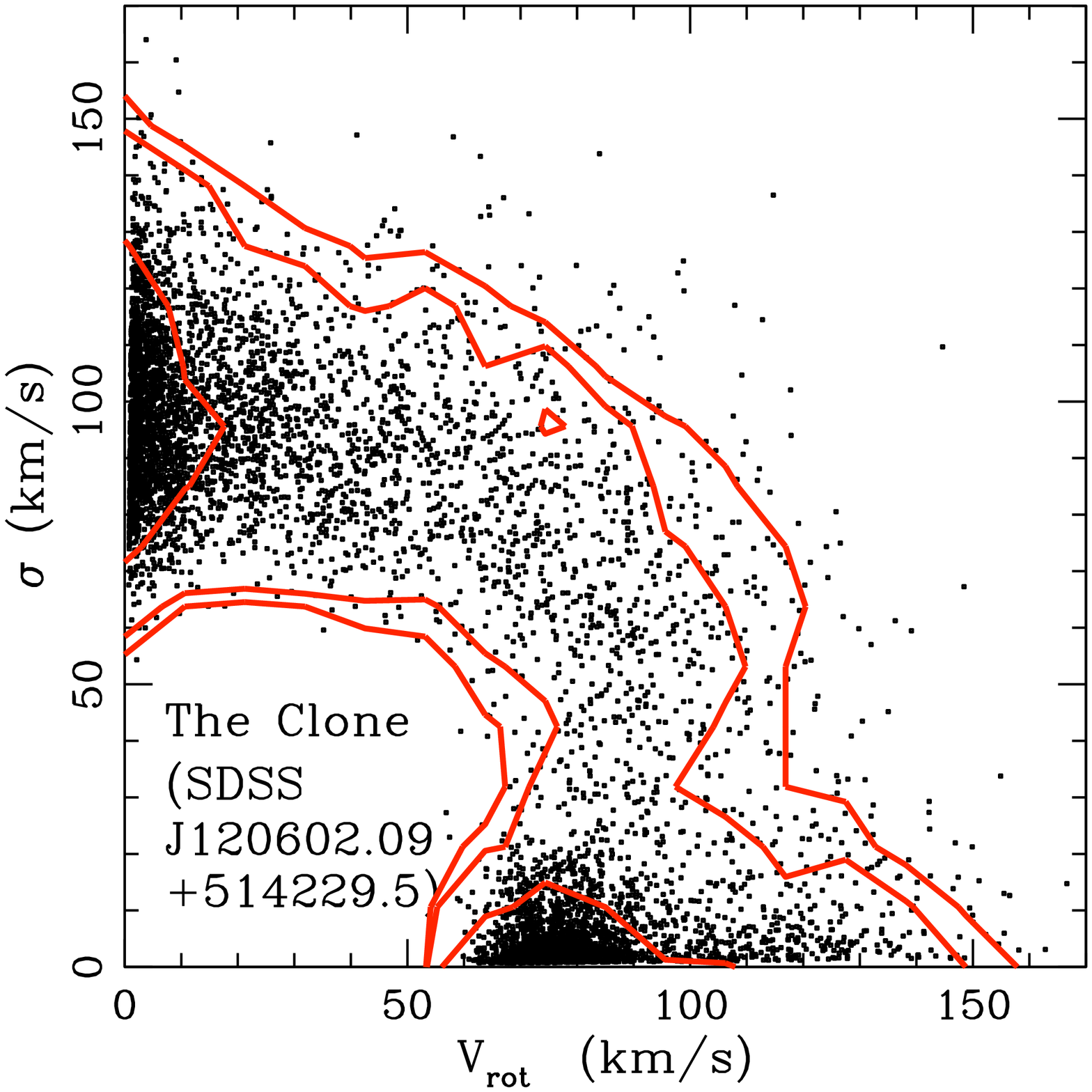}
\caption{Probability distribution for the kinematic support 
parameters $v_{rot}$ and $\sigma_g$ for \s0901full\ (left)
and for the Clone (\clonefull, right), based on our disk+halo
rotation curve model (section~\ref{sec:fitting}).  The plotted points
show nearly 8000 samples from our Markov Chain Monte Carlo parameter
fitting code.  Plotted contours enclose 68\%, 95\%, and 98\% 
confidence regions.
For \sdss0901, rotationally supported solutions are strongly favored,
with only $0.5\%$ of the samples having $\sigma_g > v_{rot}$.
For \clone, which has a lower signal-to-noise ratio, the corresponding
fraction is $\sim 44\%$. 
}
\label{fig:vr_sig}
\end{figure}

Over 99\% of the MCMC solutions for \sdss0901\ under 
the disk+halo rotation curve
model lie in a single region of parameter space, where 
rotation dominates over a small or negligible amount of random motion.
A representative fit is shown in figure~\ref{fig:s0901_spec},
and its parameters are given in table~\ref{tab:bestfitparams}.
The distribution of MCMC samples in the ($\vrot$, $\sigma_g$) plane 
is shown in figure~\ref{fig:vr_sig}.  Only 5\% of the samples
have $\sigma_g > 43.5 \kms$; 1\% have $\sigma_g > 60 \kms$;
and only 0.5\% have $\sigma_g > \vrot$.
Moreover, the rotational motions are usually dominated by the disk 
component (with $\langle \log(y_{dh}) \rangle = 0.92$, and with
$y_{dh} > 1$ in 85\% of samples).  

This conclusion is robust, being readily explained by examination of
the observed line profile.  The line is double-horned with a central
dip, and with rotation curve sides that are steep but not vertical.
Models dominated by velocity dispersion ($\sigma_g$) are expected to
be single-peaked.  Models dominated by a halo (with small core
radius and without a significant disk contribution)
would have vertical sides, except for the influence of the velocity
dispersion.  Under our adopted rotation curve model, the best way to
get a double-peaked profile with slanting sides is to have a rotation
curve that rises to a peak (driven by the disk component) and then
declines towards somewhat smaller rotation speeds (perhaps dominated
by the halo component). 

Our [CII] line profile fitting in \clone\ yields a range of viable
models.  The maximum likelihood fits are double-peaked models
with very small velocity dispersions, much like the fits for \sdss0901.  
However, these models occur in a rather small volume of the sampled
parameter space.  The MCMC fitting for \clone\ also yields single-peaked,
dispersion-dominated fits (with $y_{\sigma c} > 1$) 
in about 45\% of all samples.  These have appreciably lower 
likelihoods, but occupy a large volume in parameter space.  
For \clone, we have plotted a few different models, all statistically
acceptable fits to the data. 
These include 
the maximum likelihood model identified, which is dominated by a disk
with $\vdhalf = 78 \kms$ and $x_{ds} = 1.52$, 
and has $\sigma_g \approx 1 \kms$.
We plot also a second double-peaked profile and two dispersion-dominated
single peaked profiles.  Parameters for all are summarized in
table~\ref{tab:bestfitparams}.  The last of these plotted models was
selected to be consistent with the parameters 
that \citet{Jones10} dervied based on H$\alpha$ observations (see
section~\ref{sec:halpha}).

Our fitting yields other parameters in addition to $\vrot$ and $\sigma_g$.
The most robust and model-independent are the
redshifts (which we report in the local standard of rest [LSR] frame).  
For \sdss0901, we find
$z_{[CII]} = 2.25860 \pm 0.00007$.
This matches precisely the rest-frame optical redshift 
\citep[$z=2.2586 \pm \sim 0.0001$;][]{Hainline09}
derived from H$\alpha$, [NII], and [OIII] emission lines, 
while it is slightly higher than
the rest-UV redshift of $z=2.2558 \pm 0.0003$ \citep{Diehl09}.
This gives a net {blue}-shift of
$\Delta v \approx 260 \pm 30 \kms$ for the UV spectral features, relative
to [CII].  The \citet{Diehl09} redshift is from a cross-correlation of
the \sdss0901\ spectrum with a template from \citet{Shapley03}, making it
a weighted average of the \lya\ line redshift and the redshift from
UV absorption lines (which are, individually, mostly too faint to measure
well in \sdss0901).  To the extent that this represents a \lya\ redshift, 
it is an unusual result--- 
\lya\ is generally redshifted with respect to other velocity
tracers \citep[e.g.][]{Shapley03,McLinden11}, and a blueshift of
this magnitude for \lya\ would be worth further investigation.

For \clone, we find $z = 2.0030 \pm 0.00009$.
This can be compared to the measurements $z=2.0001 \pm 0.0006$
and $z=2.0010 \pm 0.0009$, both from rest-UV absorption lines
reported by \citet{Lin09} but based on different spectra taken at
two different telescopes.  Comparing our [CII] redshift to the
weighted average, $z_{UV} = 2.0004 \pm 0.0005$, we infer that
the gas responsible for the UV absorption lines is blue-shifted 
by $260 \pm 60 \kms$ with respect to the [CII] emitting gas.
This is typical of the absorption line blueshifts that are seen in many
high-redshift galaxies, and that are thought to be caused by 
absorption by gas in a galactic wind \citep[e.g.][]{Pettini01,
Frye02,Shapley03}.

Finally, we consider how our conclusions would be affected by changing
some of the assumptions behind our rotating disk model.  
First, we explore the possible effects of differential
gravitational magnification by applying separate amplification factors to 
the approaching and receding portions of the galaxy.  Taking the overall
mean amplification to be $\mubar$ and the ratio of red-side to
blue-side amplification to be $\mu_a$, we have amplifications of $2
\mubar / (1 + \mu_a)$ for the blue side and $2 \mu_a \mubar / (1 + \mu_a)$
for the red side.  We apply this differential magnification to the
predicted line profile before accounting for the velocity dispersion,
thus avoiding a sharp step function at the systemic velocity. 

For \sdss0901, we find that the distribution of this new parameter is
sharply peaked near unity: 98.64\% of simulated samples are in a
primary peak centered at $\mu_a = 0.92$, with dispersion of $\pm
0.17$.  The 1.3\% of simulations with more extreme ratios are all
velocity-dispersion dominated models, where the asymmetry introduced
in the magnification is hidden by the random motions of the gas.  The
only correlation between $\mu_a$ and other model
parameters is a trivial correlation with model redshift: 
An offset in redshift that would move the modeled flux centroid away 
from the observed flux centroid can be compensated by raising the peak 
of the line profile that falls nearest the observed flux centroid, 
and lowering the amplitude of the other peak.  The addition of $\mu_a$
does not alter the distributions of other model parameters in any
qualitatively important way.

For \clone, the $\mu_a$ parameter is not strongly constrained.  Models
having $\sigma_g < \vchalf$ generally have $0.4<\mu_a<2.5$, but 
dispersion-dominated models can yield acceptable fits
with any value of $\mu_a$.  With the addition of $\mu_a$, about 3/4 of
the fits end up with $\sigma_g > \vchalf$.  

We also explored the possibility of a ring-like gas distribution. 
We take the limiting case by assuming {\it all} the
emitting gas rotates with a single circular speed $v_c$.
The intrinsic line profile is then proportional to $1/\sqrt{1-(v/v_c)^2}$,
potentially modified by the differential amplification parameter $\mu_a$
described above, and convolved with a Gaussian of width $\sigma_g$ to
account for velocity dispersion within the ring.   In total, this
is a five-parameter model:
$\left\{v_c, \sigma_g, z, \fcii, \mu_a\right\}$. 
Qualitatively, our key results are unchanged under this model: 
We again find that \sdss0901\ is rotationally supported, and that
\clone\ admits a range of acceptable models.  In closer detail, we do
see some differences.  The full disk models described above can smooth the
line profile of a single ring both through velocity dispersion and
through the combination of multiple rings with different circular speeds.  
The single-ring model can only soften the
sharp edges of the double-horned profile through the effects of
velocity dispersion.   Thus, values
of $\sigma_g \la 30 \kms$ are common in the full disk model for \sdss0901\ 
but rare among the single-ring fits.  Similarly, the distribution of 
$y_{\sigma c} = \sigma_g/v_c$ tends to slightly higher values under the
single-ring model. 
Our exploration of ringlike gas distributions leads to two main conclusions.
First, ALMA observations will be tremendously valuable in 
determining the spatial distribution of the emitting gas in cases like these.
Second, our conclusion that \sdss0901\ is rotationally supported
is robust to plausible variations in the gas geometry.

\section{Comparison with H$\alpha$ kinematics}
\label{sec:halpha} 
Both \sdss0901\ and \clone\ have previously published kinematic constraints
in the form of H$\alpha$ line widths ($\Delta v_{H\alpha,obs}$).
To compare our [CII] results to these, we use our kinematic models  
to predict the H$\alpha$ line width $\Delta v_{H\alpha,pred}$ 
that would be expected, after accounting for the lower spectral resolutions 
of the published H$\alpha$ observations.  We do this by convolving our
model line profile(s) with a Gaussian whose width $\delta \lambda_{LSF}$ 
matches the reported line spread function of the H$\alpha$ observations;
measuring the FWHM $\Delta \lambda_{H\alpha,conv}$ of the convolved line 
profile; and calculating the LSF-corrected line width 
$\Delta \lambda_{H\alpha,pred} = (\Delta \lambda_{H\alpha,conv}^2
- \delta \lambda_{LSF}^2)^{0.5}$.  We finally convert the result back
to velocity units for ease of comparison ($\Delta v_{H\alpha,pred}  = \Delta \lambda_{H\alpha,pred} \times c / \lambda_{H\alpha}$).

For \sdss0901, \citet{Hainline09}  report 
$\Delta v_{H\alpha,obs} = 308 \pm 12 \kms$ (FWHM), with
a spectral resolution 
$\delta \lambda_{LSF} = 15$\AA\ FWHM,
corresponding to $210 \kms$ FWHM at $2.137 \mu m$ wavelength.
For our best-fitting model, the procedure described above
yields $\Delta v_{H\alpha,pred} = 266 \kms$.  If we apply the same
procedure to the full set of models  produced by the MCMC fitting, the
mean and standard deviation of the result becomes
$\Delta v_{H\alpha,pred} =  255 \pm 19 \kms$.  
Combining in quadrature the scatter in our simulations and the uncertainty 
reported by \citet{Hainline09},
our models predict a line width that is smaller than the observed
H$\alpha$ line width at the $2\sigma$ level.

If the difference is real, there are a few possible explanations.
First, the true LSF of the Keck spectrum could be 
18--21\AA\ (250--300$\kms$) FWHM, rather than the published measurement
of 15\AA\ (210$\kms$) FWHM \citep{Hainline09}.
Second, the [CII] and H$\alpha$ observations might sample different portions of
the galactic disk.  The {\it Herschel} observations certainly sample
the entire galaxy, while NIRSPEC's $0.76"$
slit could omit outer regions of the galactic disk.
Third, the kinematics of [CII] and H$\alpha$ could be genuinely different. 
If large velocity dispersions are associated with 
turbulence driven by star formation activity \citep{Green10,Lehnert13},
the kinematic effects of star formation would be most strongly
apparent in H$\alpha$, which is closely associated with star formation
activity.  The PDR gas that contributes strongly to [CII]
emission is found at the ``skin'' layers of molecular clouds, and may
yet retain the comparatively quiescent kinematics of molecular gas.

H$\alpha$ kinematic constraints for \clone\ have been reported
both by \citet{Hainline09} (slit spectroscopy) and
\citet{Jones10,Jones13} (integral field unit [IFU] spectroscopy). 
To compare to  \citet{Hainline09}, we followed the same procedure
we used for \sdss0901. The best fitting model gives an 
expected line width of $\Delta v_{H\alpha,pred} = 160 \kms$ FWHM, 
while the mean and 
standard deviation for all the MCMC runs was 
$\Delta v_{H\alpha,pred} = 194 \pm 52 \kms$.  
The observed H$\alpha$ line width is consistent with these 
expectations, at $\Delta v_{H\alpha,obs} = 188 \pm 9 \kms$ FWHM
(again with $\delta \lambda_{LSF} = 15$\AA\ FWHM).

The H$\alpha$ IFU spectra reported by \citet{Jones10,Jones13} offer
a more detailed comparison, with spatially resolved spectra at 
a spectral resolution of $52\kms$.  Their 
full measured rotation curve (i.e. the range of centroid
velocities of H$\alpha$ emission across the resolved image)
spans $\Delta v = 159 \pm 38 \kms$, with a
line-of-sight velocity  dispersion of $104 \pm 37\kms$
\citep[table~4]{Jones13}.  The results
depend, again, on subtracting the instrumental LSF width in quadrature,
but this time the LSF is narrower than the measured line width.
The $104\kms$ velocity dispersion corresponds directly to our $\sigma_g$
parameter.  The correspondence between the \citet{Jones13} 
parameter $\Delta v = 159 \pm 38\kms$ and our rotation speed
parameter $\vrot$ is less direct.  While we roughly expect 
$\Delta v \approx 2\vrot$, our parameter refers to the rotation
speed at the half-light radius of the [CII] emission,
while $\Delta V$ is measured across the full extent of the H$\alpha$ 
emission.  Still, we expect the kinematic parameters from
\citet{Jones13} to fall in the range $60 \kms \la \vrot\ \la 80\kms$ 
and $\sigma_g \approx 104 \kms$. 
While this is far from our {\it best-fitting}
model for the [CII] rotation curve, it is within the 90\% confidence region 
derived from the MCMC fitting results.
The MCMC approach thus provides a better consistency check than a 
simple comparison of best-fit models.

\section{Conclusions} 
\label{sec:discuss}
Our [CII] spectra of \sdss0901\ and \clone\ showcase the potential
of heterodyne spectroscopy for kinematic studies of high-redshift
galaxies.  
\citet{ForsterSchreiber09} find that about 1/3 of $z\sim 2$ galaxies
are reasonably described as rotation-dominated disks, based on
H$\alpha$ integral field unit spectra.  However, they are not able to
resolve spectral features narrower than $\sim 60\kms$.  Thus, our
measured upper limits on the velocity dispersion of \sdss0901,
$\sigma_g < 23\kms$ ($46 \kms$) at $1 \sigma$ ($2\sigma$), are
one of the tightest limits to date on the velocity dispersion of a
high redshift galaxy.  While our results are less definitive for
\clone, the maximum likelihood fits have $\sigma < 4 \kms$ ($14 \kms$)
at $1 \sigma$ ($2\sigma$), and 65\% of our simulations yield $\Delta v
/ (2 \sigma_g) \approx \vrot / \sigma_g > 0.4$--- a (semi-empirical)
criterion advocated by \citet{ForsterSchreiber09} for a disk to be
considered rotationally supported.

Spatially resolved observations at similar spectral
resolution and signal-to-noise ratio will be possible for some $z\ga 2$
galaxies using (sub)mm interferometric line observations. 
This will enable cleaner measurements of the line-of-sight velocity
dispersion.  
{\it ALMA}\/  (the Atacama Large Millimeter Array) will
be the most powerful such instrument for the foreseeable
future, offering spectral resolution comparable to HIFI, coupled
with higher sensitivity and spatial resolution as fine as $\sim 0.2''$. 
For our two particular objects, {\it ALMA}\/  will not supersede 
our {\it Herschel} [CII] observations:  The [CII] line in \sdss0901\ lies
in a gap between {\it ALMA}'s frequency bands, and \clone\ is too far north
($\delta = +51^\circ$) for {\it ALMA}\/  observation.  Fortunately, 
other strong lens systems 
will unlock the full potential of {\it ALMA}\/  for this science.

In conclusion, the {\it Herschel} observations of the [CII] 158 $\mu$m
line that we present here underscore the promise
of the bright [CII] line as a kinematic tracer for high redshift
galaxies.  They provide a unique look at the internal
dynamics of two $z\approx 2$ galaxies,
and show that very small internal velocity dispersions can be found
in the high redshift universe.

\begin{figure}
% \plottwo{example_rotcurves.pdf}{example_profiles.pdf}
\plottwo{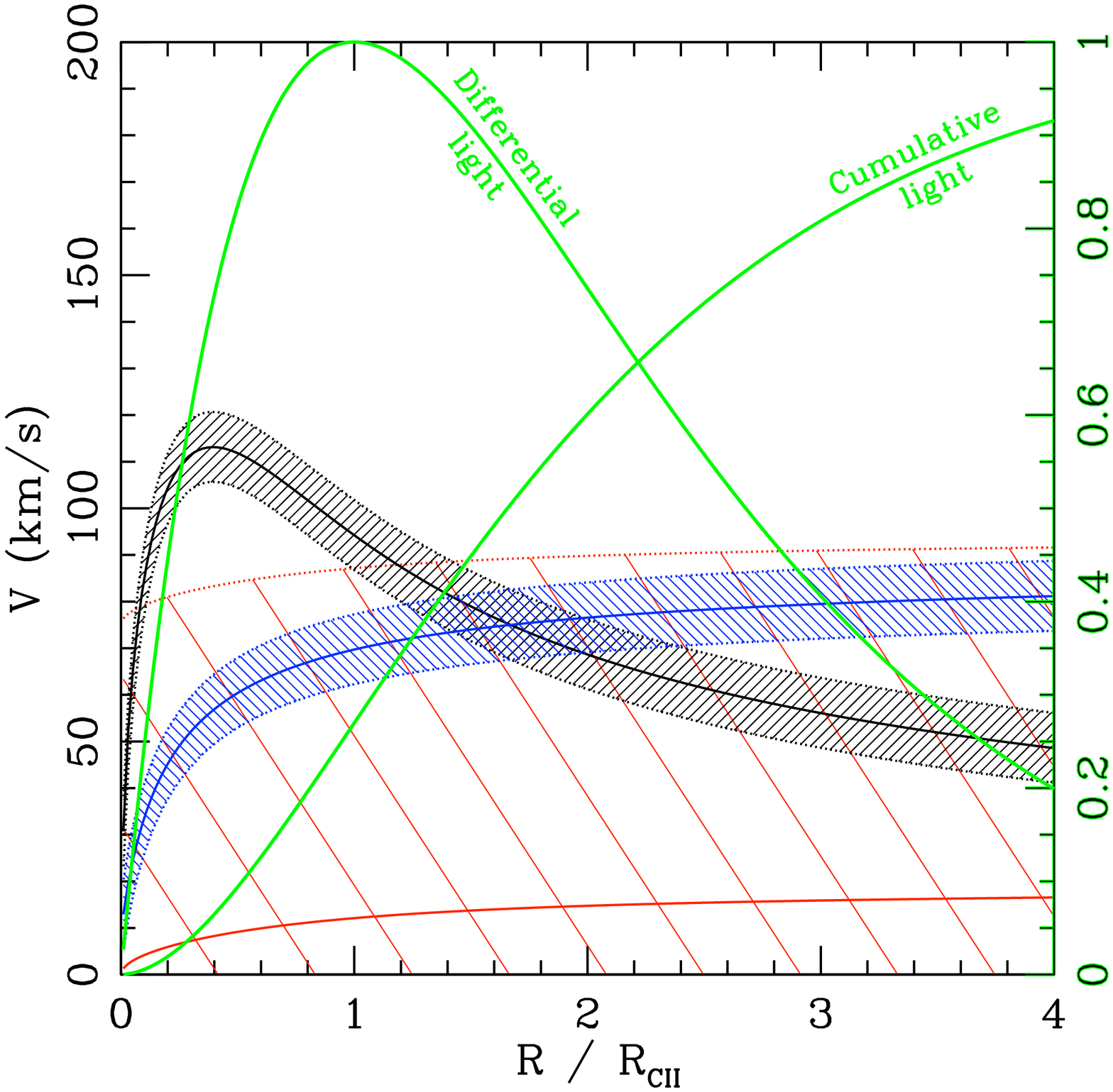}{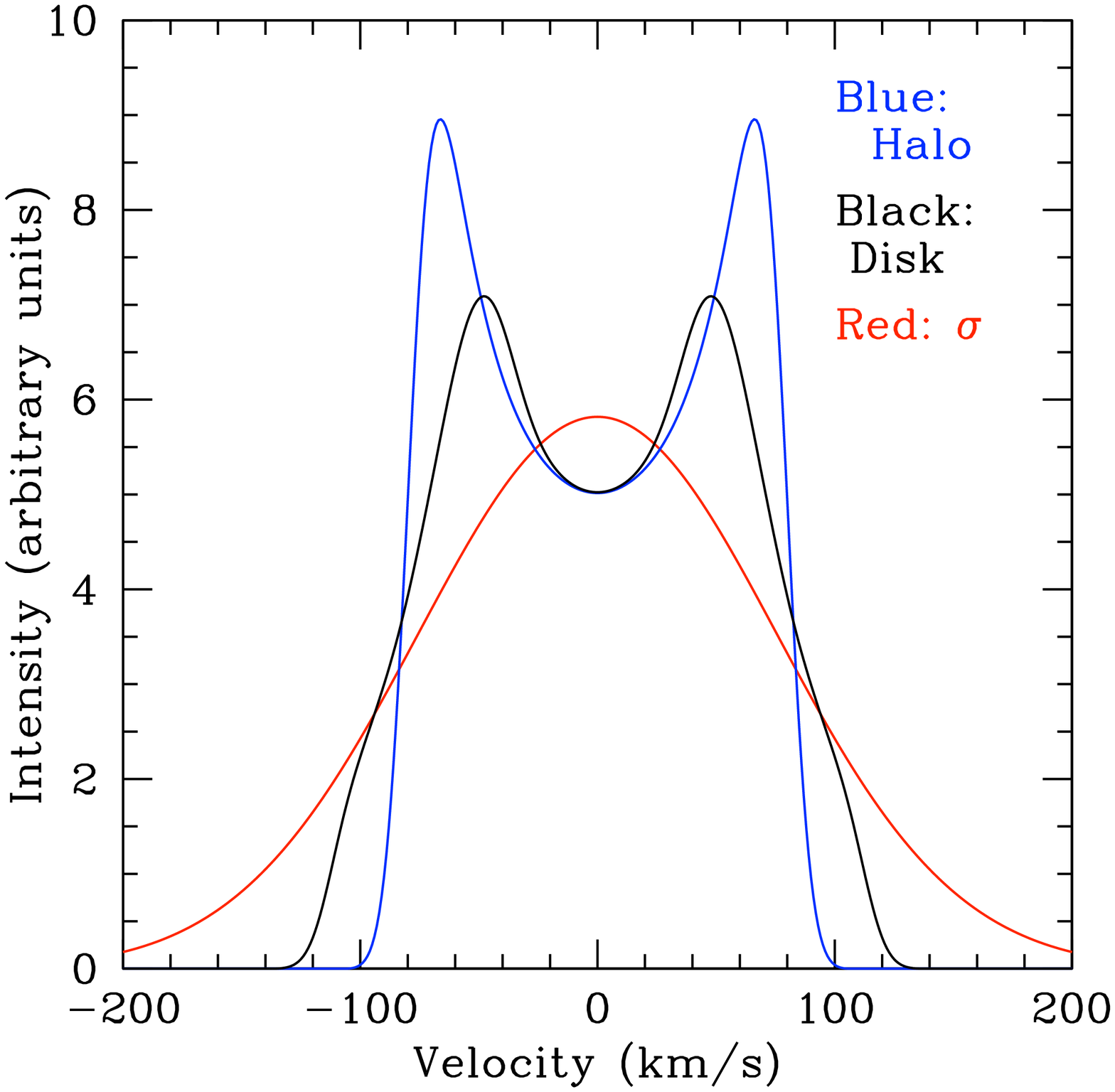}
\caption{This figure illustrates the correspondence between rotation
curve and line profile for three ``toy models,''  each of which
is dominated by a single kinematic component.  In both panels, blue corresponds
to a halo-dominated rotation curve (with gas velocity dispersion
$\sigma_g =7.5\kms$), black
to a disk-dominated rotation curve ($\sigma_g=7.5\kms$), and
red to a dispersion-dominated galaxy ($\sigma_g=75 \kms$).
{\it Left:}\/ The rotation curves (with shading for the
region $v_c(r)\pm \sigma_g$).
The two green curves show $dL_{\rm [CII]}/dr$ and its integral, $L_{\rm [CII]}(r)$,
(i.e., the enclosed [CII] luminosity), 
both normalized to their respective peaks.  They show how the different
portions of the rotation curve are weighted in the line profile.
{\it Right:}\/ The resulting line profiles. 
}
\label{fig:examples}
\end{figure}

\begin{table}
\begin{tabular}{lllllllll}
Object  & $\vdhalf$ & $\vhhalf$ & $\sigma_g$ & $x_{d}$ & $x_{h}$ &  Flux &
 $\Delta \log(L)$\tablenotemark{a} & Line\\
        &   ($\kms$)&   ($\kms$)&  ($\kms$) &           &         & ($\Kkms$) &  & style \\
\tableline
SDSS~0901 &{\bf 121} & 2 &{\bf 29} & {\bf 0.11} &  9.3  &  1.359 & 0 & 
solid red \\
\tableline
\clone    & {\bf 78} & 2 &   {\bf 1}  &  1.52  &  0.7  &  0.2657 & 0 & 
dashed red\\
\clone    & {\bf 20} & {\bf 100} & {\bf 8} & {\bf 0.07} & {\bf 6.2} & 0.365 &  -1.4 & solid red \\
\clone    &   0     & 2    & {\bf 98}  &  0.5   &  0.26 & 0.371 & -3.3 & dotted red\\
\clone    & {\bf 70} & 2 & {\bf 94} & {\bf 1.1} & 0.05 & 0.386 & -3.44 & solid blue\\
\tableline
(none)    & {\bf 75}    & 1    & 7.5  & {\bf 0.22} &  0.1  &  -  & - & black\\
(none)    &   1     & {\bf 75} & 7.5  &  1     &  {\bf 0.1}  & - & - & blue \\
(none)    &  10     & 10   & {\bf 75}  &  1     &  1    &  -  & -  & red    
\end{tabular}
\caption{Parameters of selected model fits plotted in the figures.
\label{tab:bestfitparams}}
\tablecomments{
We include the best fitting model for \sdss0901\ (top line, 
plotted in figure~\ref{fig:s0901_spec});
four representative model fits for \clone\ (lines 2--5,
plotted in figure~\ref{fig:clone_spec});
and three sample ``toy models'' chosen to illustrate features of 
the rotation curve model (lines 6--8, plotted in figure~\ref{fig:examples}).
In each case, parameters associated with the dominant component(s) determining
the line profile shape are given in bold face, and small changes to the 
remaining parameters would have small or negligible effects on the model 
line profile.
}
\tablenotetext{a}{{}$\Delta\log(L)$ is the log of likelihood, relative to the maximum likelihood
model identified for the same observed line.}
\end{table}

\section*{Acknowledgments}
We are grateful to the DARK Cosmology Centre in Copenhagen, Denmark;
Nordea Fonden in Copenhagen; the Institute for Advanced Study in 
Princeton, NJ; and Princeton University's Department of Astrophysical
Sciences for hospitality during the completion of this work.
We thank the staff at the NASA Herschel Science Center, and
Adwin Boogert in particular, for assistance in planning the observations.
We thank Mike Gladders, Casey Papovich, and Min-Su Shin
for their contributions to the
HELLO project.
We thank Scott Tremaine and Chuck Keeton for helpful discussions.
We thank an anonymous referee for their constructive suggestions.
This work has been supported by NASA through Herschel GO funding.
%

% Incorporate the contents of the bbl file here

\label{lastpage}
\end{document}